\documentclass[aps,preprint,groupeaddress,showpacs]{revtex4}
\usepackage{epsf}
\usepackage{graphicx}
\begin{document}
\title{Quasiclassical double photoionization from the $2^{1,3}S$ excited
states of helium including shakeoff}

\author{A. Emmanouilidou}
\author{T. Schneider}
\author{J.-M. Rost}
\affiliation{Max Planck Institute for the Physics of Complex Systems, N$\ddot{o}$thnitzer
Stra${\beta}$e 38, 01187 Dresden, Germany}

\begin{abstract}
We account for the different symmetries of the $2^{1,3}S$ helium excited
states in a quasiclassical description of the knockout mechanism augmented by
a quantum shakeoff contribution. We are thus
able to formulate the separate contribution of the knockout and shakeoff 
mechanisms for double photoionization for {\it{any}} excess
energy from the $2^{1,3}S$ states. Photoionization ratios and singly differential cross sections calculated
for the $2^{1,3}S$ excited states of helium are found to be in very good 
agreement with recent theoretical results.   
\end{abstract}

\vspace{0.3cm}

\maketitle

A two electron transition in an atom after the absorption of a single
photon is only possible due to electron-electron correlations. The study
of two electron transitions in helium, the simplest atomic target with two electrons, probes the 
role of electron-electron correlations in their purest form. As a result, there has
been an extensive amount of experimental and theoretical studies 
of double photoionization from the ground state of the helium
atom. 

In addition to the role of electron-electron correlations, the study of
two electron escape by a single photon absorption from the singlet, $2^{1}S$,
 and triplet, $2^{3}S$, excited states of helium probes the role of 
symmetry in the two electron
dynamics. This is particularly important for quasiclassical approximations
where symmetries and quantum interferences are difficult to include. With the exception of a few earlier studies in the high energy
limit \cite{Teng,Forrey}, it is only recently that the study of double photoionization from the $2^{1,3}S$ 
excited states of helium has attracted considerable
theoretical interest. Experimental measurements of photon absorption cross
sections from the helium excited states still remain a challenge. In these recent studies sophisticated fully numerical 
ab-initio 
methods namely the R-matrix method \cite{Hart} and the convergent 
close-coupling method \cite{Kheifets} are used to compute integral
double ionization cross sections, double to single ratios, and only 
very recently, single
and triple differential cross sections as well \cite{Colgan}. 
 However, due to the numerical nature of these sophisticated ab-initio
calculations approximate methods are also essential in uncovering the
 underlying mechanism of the two electron dynamics in the double escape
process. Such
approximate methods have been
successfully developed for the double photoionization from the helium ground
state \cite{Rost,Istomin}. The current work focuses on an approximate method to describe double
photoionization from the helium excited states.

In the double photoionization process the redistribution of the energy,
 following the photon absorption, is often discussed in terms of 
two mechanisms, that is, knockout and shakeoff dominant in the low and high
 energy limit, respectively. In the knockout mechanism one electron, the
 primary, absorbs the photon and undergoes a hard collision with the 
secondary electron, thus, knocking it out of the atom. The knockout mechanism
dominates at low energies where the interaction time of the two electrons
is large and can be described classically \cite{Rost,Schneider}. For knockout, electron correlations are essential in the final continuum state. 
 On the other hand, in the shakeoff mechanism the primary electron absorbs
the photon and leaves the atom very fast without undergoing a direct
interaction with the secondary electron. The 
secondary electron feels the sudden change in the atomic field and  
relaxes in one of 
the bound or continuum eigenstates of the remaining ion.
  The shakeoff mechanism is quantum mechanical in nature and prevails
at high energies where the interaction time of the two electrons is short.
 For shakeoff, electron correlations are important in the initial state before the photon is absorbed.
            
In the following, we formulate the explicit but separate 
contribution of both mechanisms for double photoionization from 
the $2^{1,3}S$ helium excited states for {\it any} excess energy by extending
the model developed to separate the contribution of both mechanisms for
the helium ground state in ref.\cite{Rost}.
  The separation
is achieved by a {\it quasiclassical} formulation of the knockout process.
 By construction it is free from any shakeoff contribution which is purely
 {\it quantum 
mechanical}. This separation not only facilitates the calculation of 
double photoionization but offers considerable insight into the process
concerning for example the similarity with electron impact ionization of
$He^{+}$.
 Compared to the ground state, formulating
the separate contribution of both mechanisms from the helium excited states
is a much harder problem since the different symmetry of the $2^{1,3}S$
states has to be accounted for in the model.

We express the two steps of the double photoionization 
process from the helium excited states, absorption of the photon and afterwards redistribution of 
the energy, as
\begin{equation}
\label{eq:crosssection}
\sigma^{++}_{X}=\sigma_{abs}P^{++}_{X},
\end{equation}
where X stands for either knockout or shakeoff and $P^{++}_{X}$ is 
the double electron ionization probability. In what follows we discuss
how to obtain $P_{KO}^{++}$ and then $P_{SO}^{++}$ for the $2^{1,3}S$ helium states. 
 
 -{\it Knockout mechanism}- After the photon absorption by the primary
electron, we describe the subsequent evolution of the two electrons using
 the classical-trajectory Monte Carlo (CTMC) phase space method. CTMC has 
been successfully used to describe charged particle impact
processes \cite{Abrines,Hardie,Cohen,Reinhold,Geyer}. We model the 
initial phase space of the trajectories quantum mechanically using
 a Wigner distribution \cite{Heller}. To do so, we first find the initial two-electron
wavefunction. Since electron correlations are not important in the 
initial state of the knockout mechanism, we choose our 
initial state as the independent electron wavefunction
\begin{equation}
\label{eq:wavefunction}
\Psi_{0}(\mbox{\boldmath$r$}_{1},\mbox{\boldmath$r$}_{2})=\phi_{1s}^{Z_{1s}}(\mbox{\boldmath$r$}_{1})\phi_{2s}^{Z_{2s}}(\mbox{\boldmath$r$}_{2}),
\end{equation}
 where $\phi_{1s}^{Z_{1s}}(\mbox{\boldmath$r$}_{1})=(Z_{1s}^{3}/\pi)^{1/2}e^{-Z_{1s}r_{1}}$
and $\phi_{2s}^{Z_{2s}}(\mbox{\boldmath$r$}_{2})=(Z_{2s}^{3}/(8\pi))^{1/2}(1-Z_{2s}r_{2}/2)e^{-Z_{2s}r_{2}/2}$
are hydrogenic $1s$ and $2s$ orbitals with effective charges $Z_{1s}$ 
and $Z_{2s}$,
respectively. $\Psi_{0}(\mbox{\boldmath$r$}_{1},\mbox{\boldmath$r$}_{2})$
given by 
Eq.(\ref{eq:wavefunction}) is not a symmetrized wavefunction, thus, we do not yet account
for the different symmetries of the triplet $2^{3}S$ and the singlet
 $2^{1}S$ helium excited states. To find the effective 
charges $Z_{1s}$ and $Z_{2s}$ we use the two-electron ionization
energies $I_{2^{1}S/2^{3}S}^{++} \approx 2.146/2.175$ for the $2^{1}S/2^{3}S$ states 
 given in ref.\cite{Forrey}, and the ionization energies
of the $1s$ and $2s$ orbitals as follows. The energy needed to remove the $1s$
electron after the $2s$ electron has been ionized is $2$ a.u. and thus
from $I_{1s}=Z_{1s}^2/2=2$ we find $Z_{1s}=2$ for both the triplet
and the singlet states. From $I_{2^{1}S/2^{3}S}^{++}-I_{1s}=I_{2s}=Z_{2s}^{2}/8$
we find $Z_{2s}\approx 1.081/1.184$ for the $2^{1}S/2^{3}S$ states, respectively. 
 Atomic units are used throughout this paper unless otherwise indicated.
 
While, for the $1s^{2}$ helium ground state 
the photon is necessarily absorbed by a $1s$ electron, 
 for the $1s2s$ configuration of the $2^{1,3}S$ helium states 
the photon can be absorbed by a $1s$ or a $2s$ electron.
 For the values of the effective charges given above,
 one can show that the cross section is much larger for photon absorption
from a $1s$ electron rather than a $2s$ electron using the independent
electron picture \cite{Amusia}. Therefore, we take the primary
electron to be the one on the $1s$ orbital.
 With the PEAK approximation, exact in the high photon energy limit \cite{Kabir}, we assume that the photon absorption happens
directly at the nucleus. This significantly reduces the
initial phase space to be sampled by the CTMC method.

 The initial phase space distribution, $\rho(\mbox{\boldmath$\Gamma$})$, is the two electron 
density immediately after the photon absorption given by 
\begin{equation}
\label{eq:distribution}
\rho(\mbox{\boldmath$\Gamma$})=N \delta(\mbox{\boldmath$r$}_{1})
\rho_{2}(\mbox{\boldmath$r$}_{2},\mbox{\boldmath$p$}_{2}),  
\end{equation}
where N is a normalization constant. The initial distribution of the primary
electron is $\delta(\mbox{\boldmath$r_{1}$})$ (PEAK approximation)
 while that of the secondary electron is given by
\begin{equation}
\label{eq:distribution2}
\rho_{2}(\mbox{\boldmath$r$}_{2},\mbox{\boldmath$p$}_{2})=
W_{\psi}(\mbox{\boldmath$r$}_{2},\mbox{\boldmath$p$}_{2})\delta(\epsilon_{2}^{in}
-\epsilon_{B}),
\end{equation}   
where $W_{\psi}(\mbox{\boldmath$r$}_{2},\mbox{\boldmath$p$}_{2})$
is the Wigner distribution function of the two electron wavefunction with the 
primary electron at the nucleus, $\mbox{\boldmath$r$}_{1}=0$,
\begin{equation}
\label{eq:orbital}
\psi(\mbox{\boldmath$r$}_{2})=\Psi_{0}(\mbox{\boldmath$r$}_{1}=0,\mbox{\boldmath$r$}_{2})(\langle\Psi_{0}(\mbox{\boldmath$r$}_{1}=0,\mbox{\boldmath$r$}_{2})|\Psi_{0}(\mbox{\boldmath$r$}_{1}=0,\mbox{\boldmath$r$}_{2})\rangle)^{-1/2}. 
\end{equation}
In Eq.(\ref{eq:distribution2}), we take the energy of the secondary electron
immediately after photon absorption, $\epsilon_{2}^{in}$, to be fixed on the
$2s$ energy shell $\epsilon_{B}=-Z_{2s}^2/8$. From Eq.(\ref{eq:wavefunction}),
 it follows that $\epsilon_{2}^{in}=p_{2}^{2}/2-Z_{2s}/r_{2}$.
 The excess energy available to the two electron system after photon
absorption is determined by the photon energy to be  
\begin{equation}
\label{energy}
\begin{array}{ccc}
E=\omega-I_{2^{1}S}^{++} , & & E=\omega-I_{2^{3}S}^{++} \end{array}
\end{equation}  
for the $2^{1}S$ and $2^{3}S$ states, respectively. Due to the PEAK
approximation the primary electron can have any energy necessary so that
together with the initial energy $\epsilon_{2}^{in}$ of the secondary
electron it adds up to the excess energy E in Eq.(\ref{energy}).
 After modelling the initial phase space distribution, we propagate
the electron trajectories using the classical equations of motion (CTMC).
 Regularized coordinates \cite{regularized} are used for the propagation of
the electron trajectories to avoid problems with trajectories starting at the
nucleus ($\mbox{\boldmath$r$}_{1}=0$). Doubly ionized are those
 trajectories that end with the asymptotic energies of both electrons 
being positive. To evaluate the double electron escape probability
each trajectory is weighted by the initial phase space
 Wigner distribution.

 So far we have treated the
two electrons as distinguishable particles, that is, we distinguish between the
primary and the secondary electron. To account for the singlet and
triplet symmetries we have to symmetrize the probability 
{\it amplitude} (differential probability) $dP_{KO}^{++}/d\epsilon$
with respect to the two identical particles,
\begin{equation}
\label{eq:amplitude}
\frac{dP_{KO}^{++}}{d\epsilon}=\frac{1}{2}\left(\sqrt{\frac{dP_{KO}^{++}(\epsilon,E)}{d\epsilon}}\pm 
      \sqrt{\frac{dP_{KO}^{++}(E-\epsilon,E)}{d\epsilon}}\right)^2.
\end{equation}
 In Eq.(\ref{eq:amplitude}), $dP_{KO}^{++}(\epsilon,E)/d\epsilon$ is the probability
for both electrons to escape when the primary electron is ejected with energy
$\epsilon$, where $0\leq\epsilon \leq E$, and the secondary electron
is ejected with energy $E-\epsilon$. To evaluate
$dP_{KO}^{++}(\epsilon,E)/d\epsilon$ we divide the energy interval 
$[0,\epsilon]$ into N equally sized bins and find the doubly ionized trajectories  
which fall into the bins. In our calculations we take $N=21$ for excess
energies up to $80$ eV  and $N=27$ for higher excess energies. The double ionization
 probability $P_{KO}^{++}$ is obtained by integrating over all possible 
energies that an electron can be ejected with, that is
\begin{equation}
\label{eq:double}
P_{KO}^{++}=\int_{0}^{E}\frac{dP_{KO}^{++}}{d\epsilon}d\epsilon.
\end{equation}
Note that for the case of the helium ground state the double ionization
  probability $P_{KO}^{++}$ is worked out without using the 
differential probabilities \cite{Rost}.

-{\it Shakeoff mechanism}- Assuming that the primary electron is suddenly 
removed from the atom, Aberg \cite{Aberg} found that the probability for 
the shaken (secondary) electron to relax on a hydrogenic eigenstate of the
remaining ion for {\it {any}} excess energy is
\begin{equation}
\label{eq:shakeoff}
P_{\alpha}^{\nu}=|\langle\phi_{\alpha}|\psi^{\nu}\rangle|^{2}/\langle\psi^{\nu}|\phi^{\nu}\rangle,
\end{equation}
where $\psi^{\nu}(\mbox{\boldmath$r$}_{2})=\int d^{3} \mbox{\boldmath$r$}_{1}
\nu^{*}(\mbox{\boldmath$r$}_{1})\Psi_{0}(\mbox{\boldmath$r$}_{1},\mbox{\boldmath$r$}_{2})$
and $\nu(\mbox{\boldmath$r$}_{1})$ is
the primary electron wavefunction after it has left the atom. The primary electron is in an $s$ state
before the photon absorption and in a $p$ state
afterwards. $\Psi_{0}(\mbox{\boldmath$r$}_{1},\mbox{\boldmath$r$}_{2})$ is the
initial state wavefuntion of the $2^{1,3}S$ helium excited
states and $\phi_{\alpha}$ is a hydrogenic eigenstate of the bare nucleus ($Z=2$)
that is either a bound ($\alpha=n$) or a continuum state
($\alpha=\epsilon$). Aberg \cite{Aberg} has further shown that when the primary
electron leaves the atom with very high energy
($\nu(\mbox{\boldmath$r_{1}$})=(2\pi)^{-3/2}e^{-i{}\mbox{\boldmath$k$}_{1}{} 
\mbox{\boldmath$r$}_{1}}$) Eq.(\ref{eq:shakeoff}) takes the simplified form

\begin{equation}
\label{eq:shakeoff1}
P_{\alpha}=\frac{|\langle\phi_{\alpha}|\Psi_{0}(\mbox{\boldmath$r$}_{1}=0,\mbox{\boldmath$r$}_{2})\rangle|^{2}}{\langle\Psi_{0}(\mbox{\boldmath$r$}_{1}=0,\mbox{\boldmath$r$}_{2})|\Psi_{0}(\mbox{\boldmath$r$}_{1}=0,\mbox{\boldmath$r$}_{2})\rangle}.
\end{equation}
Eq.(\ref{eq:shakeoff1}) reveals the quantum mechanical nature of the
shakeoff process since it is expressed as an overlap of the initial bound
state wavefunction and the final continuum state wavefunction. 

 Although Eq.(\ref{eq:shakeoff1}) was derived in the high energy limit
we assume that the primary electron absorbs the photon on the nucleus 
for all excess energies, that is, we adopt the PEAK approximation as in the knockout
case. To find the double escape probability $P_{SO}^{++}$ we then integrate over all
possible energies of the shaken electron in the continuum

\begin{equation}
\label{eq:shakeoff2}
P_{SO}^{++}(E)=\int_{0}^{E} P_{\epsilon} d\epsilon.
\end{equation}  

We further simplify the evaluation of the shakeoff probability in
Eqs.(\ref{eq:shakeoff1}), (\ref{eq:shakeoff2}) by taking the 
initial state to be the symmetrized wavefunction
\begin{equation}
\label{eq:wavefunction1}
\Psi_{0}(\mbox{\boldmath$r$}_{1},\mbox{\boldmath$r$}_{2})=N_{1}\left(\phi_{1s}^{Z_{SO}^{1}}(\mbox{\boldmath$r$}_{1})\phi_{2s}^{Z_{SO}^{2}}(\mbox{\boldmath$r$}_{2})\pm  \phi_{1s}^{Z_{SO}^{1}}(\mbox{\boldmath$r$}_{2})\phi_{2s}^{Z_{SO}^{2}}(\mbox{\boldmath$r$}_{1})\right)
\end{equation}
for the singlet and triplet states, respectively, with $N_{1}$ a
normalization constant. The initial state
correlations are accounted for only through the effective charges.
We next assign the same set of effective charges $Z_{SO}^{1}$ and $Z_{SO}^{2}$
for both the triplet and the singlet states   
 as follows. The asymptotic ratio (high energy limit) of double to single 
ionization is found very accurately in ref.\cite{Forrey} to be 
$R_{\infty}=0.009033/0.003118$ for the singlet/triplet states where in our
model $R_{\infty}$ is given by
\begin{equation}
\label{eq:R}
R_{\infty}=P_{SO}^{++}(E\rightarrow \infty)/(1-P_{SO}^{++}(E\rightarrow \infty)),
\end{equation}
and
\begin{equation}
\label{eq:shakeoff4}
P_{SO}^{++}(E\rightarrow \infty)
=\int_{0}^{\infty}P_{\epsilon}d\epsilon=1-\sum_{n}P_{n}.
\end{equation}
Using Eqs.(\ref{eq:R}), (\ref{eq:shakeoff4}) and the symmetrized 
wavefunctions  
in Eq.(\ref{eq:wavefunction1}) we find the sets of charges 
that match both 
asymptotic ratios $R_{\infty}$ for the singlet and the triplet
states. We then select that set of charges for which 
the shakeoff double ionization probability as a function of
the excess energy, obtained using the simple wavefunctions given in 
Eq.(\ref{eq:wavefunction1}), is closest to the one obtained using the fully 
correlated Hylleraas wavefunctions given in ref.\cite{Huang}. The set 
of charges
thus found is $Z_{SO}^{1}\approx 1.757$ and $Z_{SO}^{2}\approx 1.728$.
 The reason we do not use the Hylleraas wavefunctions given in 
ref.\cite{Huang} is that they do not reproduce the accurate
asymptotic ratios obtained in ref.\cite{Forrey} using highly accurate 
Pekeris-type wavefunctions. We emphasise though that one does not need
to use the approximate wavefunctions in Eq.(\ref{eq:wavefunction1})
to compute the double ionization probability; highly accurate wavefunctions
that reproduce the correct asymptotic ratios could be used instead.      
 
For the shakeoff probability, $P_{\epsilon}$ in Eq.(\ref{eq:shakeoff1}) 
already gives the differential double ionization probability. Despite the
symmetrization in Eq.(\ref{eq:wavefunction1}) we have lost the
indistinguishability of the electrons by identifying one electron as the
primary one which absorbs the photon. Thus we need to symmetrize again in the
final state with respect to the 
equal energy sharing point $\epsilon=E-\epsilon=E/2$ \cite{Rost}. That is,
\begin{equation}
\label{eq:diffSO}
\frac{dP_{SO}^{++}}{d\epsilon}=\frac{1}{2}\left(\frac{dP_{\epsilon}}{d\epsilon}+
\frac{dP_{E-\epsilon}}{d\epsilon}\right).
\end{equation}     

\begin{figure}
\begin{centering}
\leavevmode
\epsfxsize=0.5\linewidth
\epsfbox{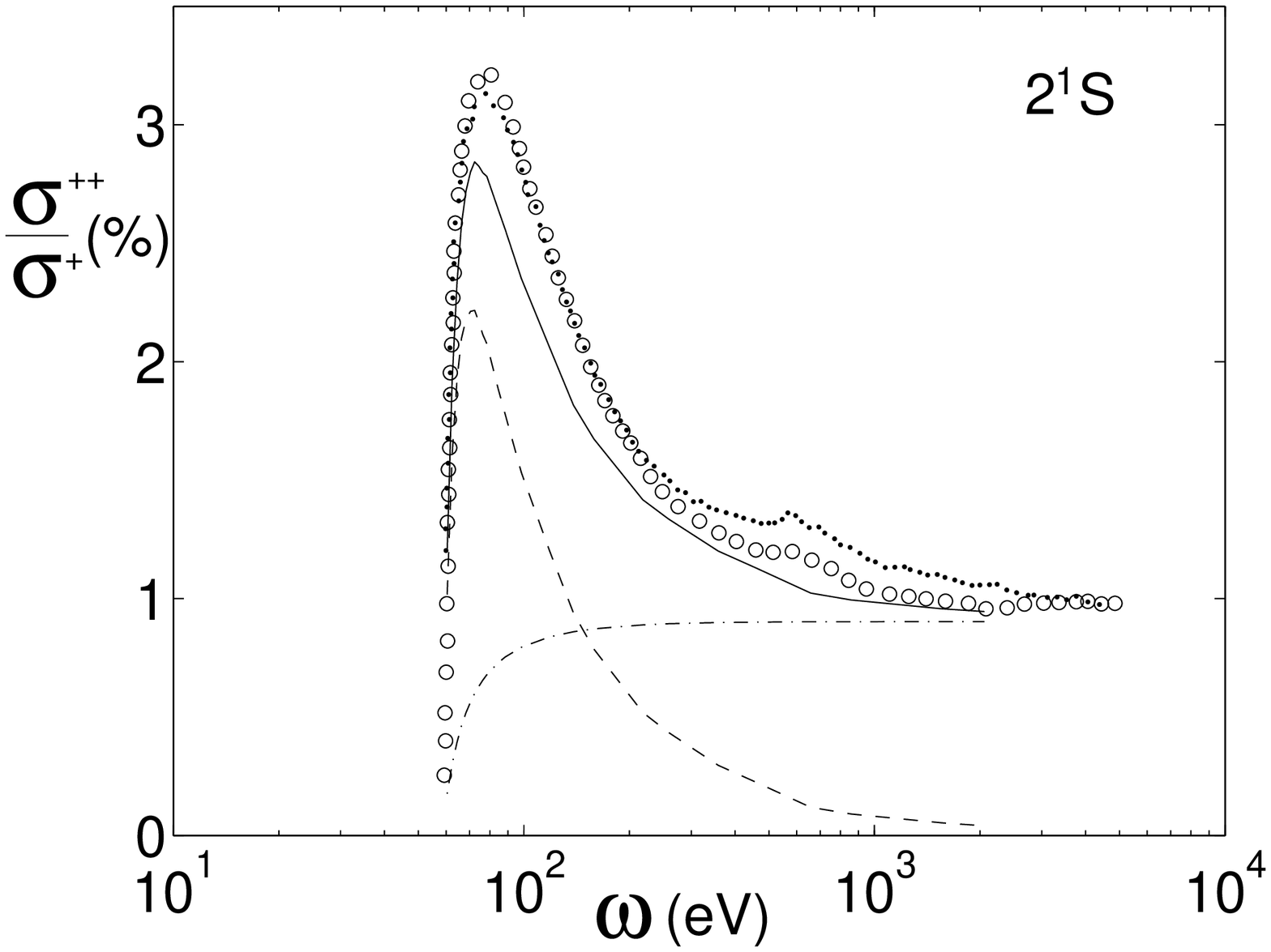}
\epsfxsize=0.5\linewidth
\epsfbox{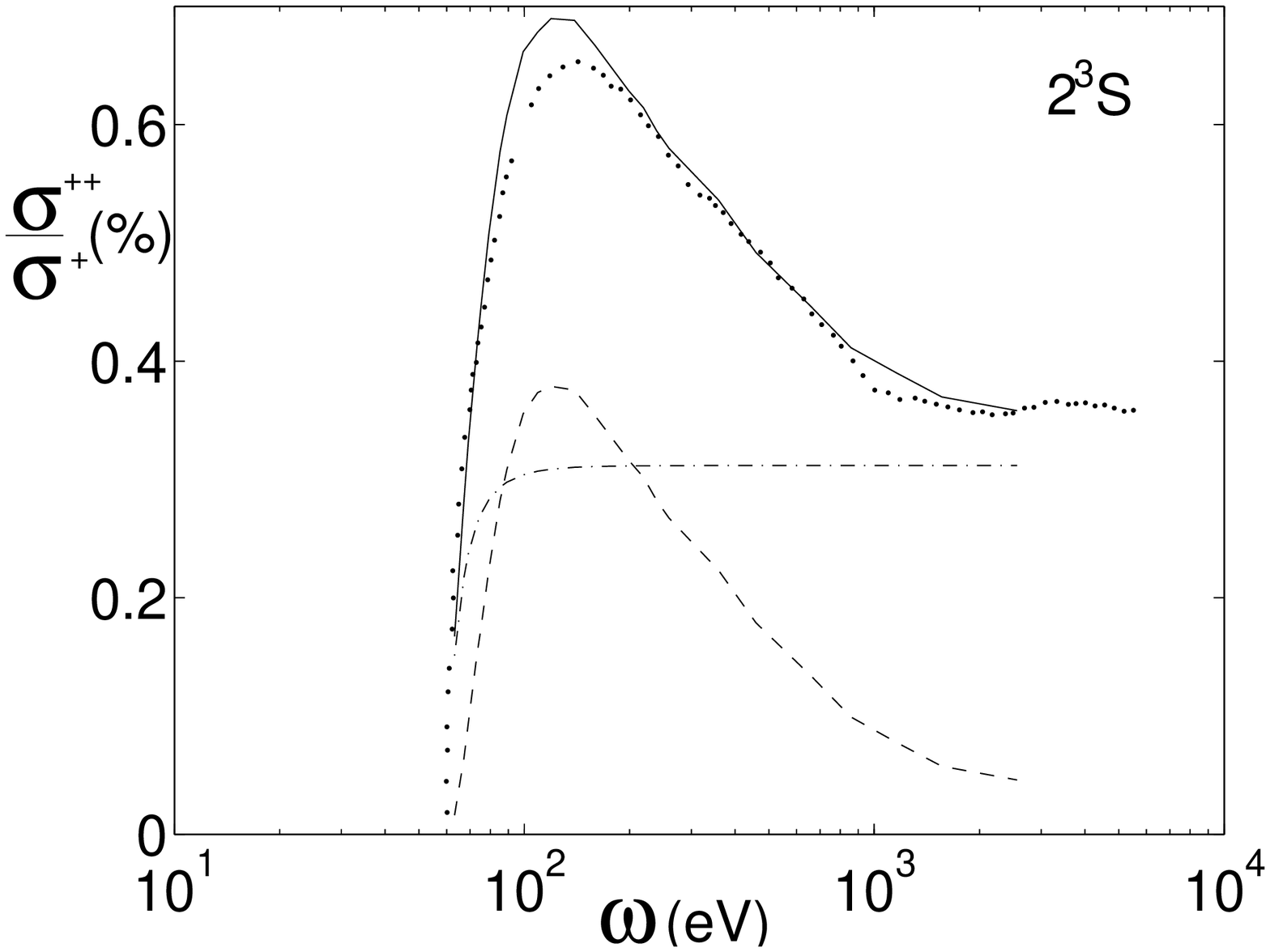}
\caption{}
\label{fig:total}
\end{centering}
\end{figure}

-{\it Photoionization ratios}-. According to Eq.(\ref{eq:crosssection}) 
$\sigma^{++}=\sigma_{abs}(P_{KO}^{++}+P_{SO}^{++})$ and
$\sigma^{+}=\sigma_{abs}-\sigma^{++}$. Thus, the double to single ionization
ratio is given by
\begin{equation}
\label{eq:ratio}
 \sigma^{++}/\sigma^{+}=P^{++}/(1-P^{++}),
\end{equation}
 where
$P^{++}=P_{KO}^{++}+P_{SO}^{++}$.
  In figure (\ref{fig:total}) we compare the double to single ratio for the
$2^{1,3}S$ helium excited states with the
results obtained by Kheifets {\it {et al}} \cite{Kheifets} using the convergent
close-coupling method and show that there
is a very good agreement. The agreement is better for the $2^{3}S$ state. 
We find that the deviation occurs, particularly for the $2^{3}S$ state,
 at photon energies where the contribution of knockout and shakeoff mechanisms
is comparable. At these energies any interference effect between the knockout
and shakeoff mechanism would have its largest effect. So, it may be that the
deviations we see are due to that interference effect that we do not account
for in our calculation, since we add the knockout and shakeoff contributions 
incoherently. 
 For the $2^{1}S$ state a maximum of $\approx$ 2.84 \%  
is reached at $14$ eV above the double ionization threshold of the $2^{1}S$
state. For the $2^{3}S$ state a maximum of $\approx$ 0.69 \% is reached at
$60$ eV above the ionization threshold of the $2^{3}S$ state. In
figure (\ref{fig:total}) we see that at high energies
the knockout contribution goes to zero as expected and the shakeoff contribution
dominates and reaches the asymptotic limit of $0.009033/0.003118$ for the
singlet/triplet states. Stronger correlation effects for the singlet symmetry
($\mbox{\boldmath $r$}_{1}=\mbox{\boldmath $r$}_{2}$ is not forbidden as is
the case for the triplet) result in a much higher double to single ionization
ratio compared to the triplet case. From figure (\ref{fig:total}), for the
$2^{1,3}S$ states, and ref.\cite{Rost}, for the helium ground state $^{1}S$, we see
that as we go from $^{1}S\rightarrow 2^{1}S \rightarrow 2^{3}S$ the shakeoff
mechanism overtakes the knockout mechanism at smaller energies. The reason is that 
 the electron-electron
correlation becomes smaller, thus diminishing the knockout contribution and 
favouring the shakeoff mechanism at even smaller energies.
 
\begin{figure}
\begin{centering}
\leavevmode
\epsfxsize=0.5\linewidth
\epsfbox{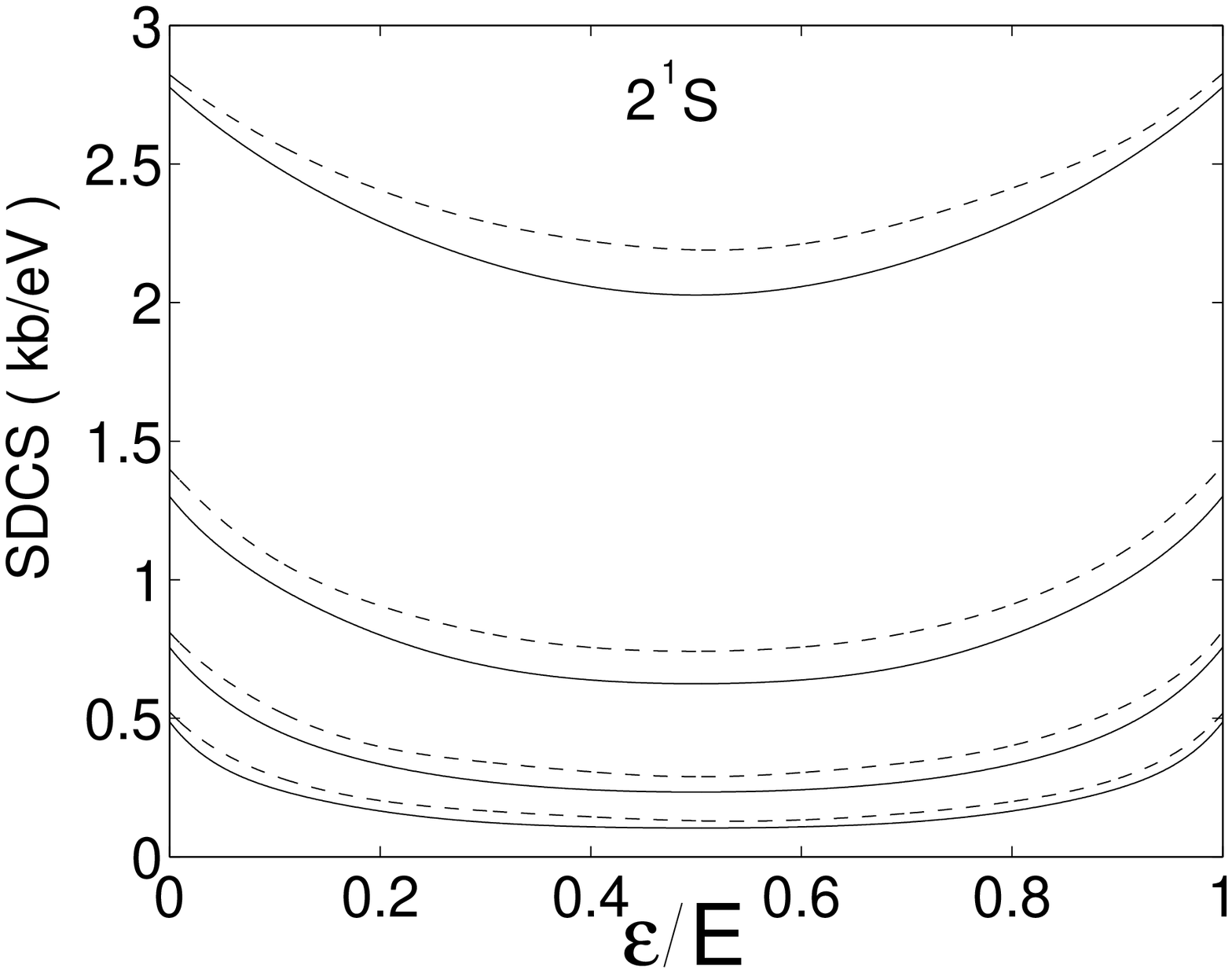}
\epsfxsize=0.5\linewidth
\epsfbox{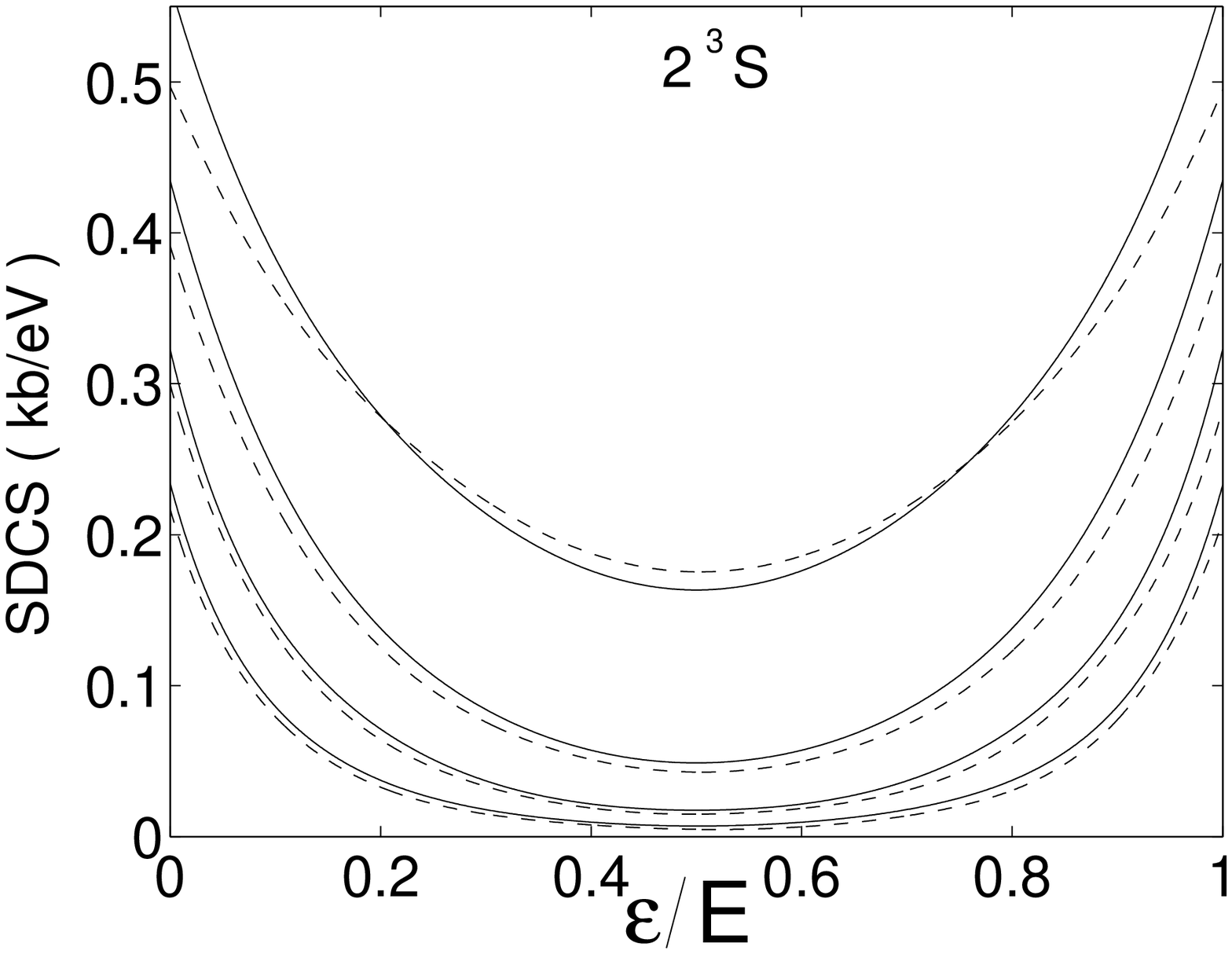}
\caption{}
\label{fig:differential}
\end{centering}
\end{figure}

-{\it Single differentials}- To compute the single differential
probabilities for the helium excited states we use
\begin{equation}
\label{eq:diffprob}  
\frac{dP^{++}}{d\epsilon}=\frac{dP_{KO}^{++}}{d\epsilon}+\frac{dP_{SO}^{++}}
{d\epsilon}.
\end{equation}   
In addition we compute the single differential cross sections using
\begin{equation}
\label{eq:diffscat}
\frac{d\sigma^{++}}{d\epsilon}=\sigma_{abs}\left(\frac{dP_{KO}^{++}}{d\epsilon}+\frac{dP_{SO}^{++}}
{d\epsilon}\right),   
\end{equation}
where for $\sigma_{abs}$ we use the results for the total-photoionization
 cross section for the $2^{1}S$ and the $2^{3}S$ states given in 
ref.\cite{Kheifets}. In figure (\ref{fig:differential}) we compare our results
for $d\sigma^{++}/d\epsilon$ with the results obtained very recently
by Colgan {\it et al} \cite{Colgan} using the convergent close-coupling
 method for four values of the excess energy. We see that our results for the
 single differential cross section as a function of the ejected electron energy
 normalized by the excess energy
are smaller for the $2^{1}S$ state while
there is an excellent agreement for the $2^{3}S$ state. Again as we go from
$^{1}S\rightarrow 2^{1}S \rightarrow 2^{3}S$ the single differential cross
 sections are more U-shaped for the same excess energy. The reason is
again that the electron-electron correlation decreases thus favouring
the ejection of one fast and one slow electron, that is favouring unequal
 energy sharing.

  In figure (\ref{fig:KOSO}) we show the separate contribution of knockout and
shakeoff to the single differential probabilities for the $2^{1,3}S$ helium
states for excess energies $10$ eV, $40$ eV and $160$ eV. For the singlet state the 
knockout contribution dominates at small excess energies, $10$ eV and $40$ eV,   
while as the excess energy is increased to $160$ eV the shakeoff contribution 
begins to 
dominate regions of unequal energy sharing. For the triplet case the shakeoff
mechanism is
already significant at small excess energies. Note, that the knockout
contribution for the triplet case is zero at the equal energy sharing point, 
$\epsilon=E-\epsilon$, because of the symmetrization with respect to the two identical
electrons, see Eq.(\ref{eq:amplitude}). 
 
In conclusion we have shown that the double ionization from the $2^{1,3}S$
states can be accurately described by a separate formulation and calculation
of the knockout and shakeoff mechanism at any excess energy. In comparison
to the helium ground state \cite{Rost} this is a harder problem because   
we have to account for the different symmetries of the singlet and
triplet states. The success of this simple model to describe double
ionization from the helium ground state as well as the helium excited states
is proof for its validity. In the future, we plan to use this simple model
to describe triple photoionization cross sections.

The authors wish to thank T. Pattard for helpful discussions.

\newpage
\begin{center}
List of Figures
\end{center}
\begin{itemize}
\item Figure 1. Double to single ionization ratio as a function of the photon
energy. Dots/open circles indicate the results of Kheifets {\it et
al} \cite{Kheifets} in the velocity/acceleration gauge. For the triplet state
Kheifets results in both gauges are indistinguishable to the scale of the figure. Our results are 
indicated by a solid line for the total, by a dashed line for the knockout and 
by a dashed-dot line for the shakeoff double to single ratio.

\item Figure 2. Absolute single differential cross sections as a function of the 
electron ejected energy scaled by the excess energy. The dashed lines are the
results by Colgan {\it et al}. Our results are indicated by solid lines.

\item Figure 3. Absolute single differential probabilities as a function of the
ejected electron energy. The knockout contribution is indicated by dashed
lines while the shakeoff by solid lines. The top panel is for the $2^{1}S$
state while the bottom is for the $2^{3}S$ state.
\end{itemize}

\newpage

\begin{figure}
\begin{centering}
\resizebox*{0.6\textwidth}{!}{\includegraphics{fig3.eps}}
\caption{}
\label{fig:KOSO}
\end{centering}
\end{figure}

\end{document}